\documentclass[aps,prl,preprint,superscriptaddress,showpacs]{revtex4}

\usepackage{graphicx,graphics,color}
\usepackage{amssymb,amsfonts,amsmath}

\begin{document}
\baselineskip 14pt
\topmargin= -0.65in
\footskip = 22pt

\title{Comment on ``Controllability of Complex Networks with Nonlinear Dynamics"}

\author{Jie Sun}
\affiliation{\small $\mbox{Department of Physics and Astronomy, Northwestern University, Evanston, IL 60208, USA}$}

\author{Sean P. Cornelius}
\affiliation{\small $\mbox{Department of Physics and Astronomy, Northwestern University, Evanston, IL 60208, USA}$}

\author{William L. Kath}
\affiliation{\small 
                       $\mbox{Engineering Sciences and Applied Mathematics, Northwestern University, Evanston, IL 60208, USA}$}
\affiliation{\small $\mbox{Northwestern Institute on Complex Systems, Northwestern University, Evanston, IL 60208, USA}$\\ {~}} 

\author{Adilson E. Motter}
\affiliation{\small $\mbox{Department of Physics and Astronomy, Northwestern University, Evanston, IL 60208, USA}$}
\affiliation{\small $\mbox{Northwestern Institute on Complex Systems, Northwestern University, Evanston, IL 60208, USA}$\\ {~}}

\begin{abstract} \baselineskip 14pt
\noindent
{~}\\
The recent paper by W.-X.\ Wang, Y.-C.\ Lai, J.\ Ren,  B.\ Li \& C.\ Grebogi [arXiv:1107.2177v1]~\cite{Wang2011}~proposed a method for the control of complex networks with nonlinear dynamics based on linearizing the system around a finite number of local desired states. The authors purport that any bidirectional network  with one-dimensional intrinsic node dynamics can be fully controlled by a single driver node, which can be any node in the network, regardless of the network topology.  According to this result, a network with an arbitrarily large number of nodes (a million, a billion, or even more) could be controlled by a single node. Here we show, however, that this result is specious and that it does not hold true even for networks with only two nodes.  We demonstrate that the erroneous results are a consequence of a  fundamental flaw in the proposed method, namely, that reaching a local desired state using the proposed linearization procedure generally requires a nonlocal trajectory, which grossly violates the linear approximation. Their further conclusion that network systems with nonlinear dynamics are more controllable than those with linear dynamics  is a known fact presented as novel based on a flawed argument.  A central problem underlying the authors' argument is that their formulation is entirely based on the number of driver nodes required to reach a desired state---even when this number is  one, keeping the system at that desired state generally requires directly controlling all nodes in the network.  When this is taken into account, the conclusion that nonlinear dynamics can facilitate network control was already anticipated in S.P.\ Cornelius, W.L.\ Kath \& A.E.\ Motter [arXiv:1105.3726v1] \cite{Cornelius2011}, which is not referenced in their paper. If one insists on using the authors' formulation, then, in contradiction to their claim, nonlinear systems would never be more controllable than linear ones simply because the controllability of linear systems requires a single driver node, as shown by N.J.\ Cowan, E.J.\ Chastain, D.A.\ Vilhena \& C.T.\ Bergstrom [arXiv:1106.2573v1] \cite{Cowan2011v1}, which is also not referenced in their paper.  Taken together, all results in Ref.\ \cite{Wang2011} are unsound, including the authors' generalizations to high-dimensional node dynamics and directed networks.
\end{abstract}

\maketitle
\baselineskip 14pt

\noindent
Controllability of nonlinear systems (including networked systems) has long been an important open problem in control theory \cite{Haynes1970,Sussmann1972,Hermann1977}.  Surprisingly strong results were recently announced by Wang {\it et al.}~in Ref.\ \cite{Wang2011}. These include a general method to control networks of coupled nonlinear dynamical systems, and the full characterization of the minimum set of driver nodes. It is proposed therein that any network with one-dimensional node dynamics can be fully controlled by the roots of any minimum set of directed trees spanning the network, which implies a single driver node in the case of undirected networks (the extension to high dimension is immediate \cite{Wang2011}). Here, we first show that the conclusions in Ref.\ \cite{Wang2011} are generally not valid. Next, we identify and discuss the main flaw that renders the proposed control method invalid. We then show that the assertion in Ref.\ \cite{Wang2011} that networks with nonlinear dynamics are more controllable than networks with linear dynamics is both not novel and based on wrong reasoning.  We also elaborate on the fact that the full control of any dynamical system, whether a network or not, concerns much more than merely driving it to a target state---it also concerns the stabilization of the target state, which in general requires more driver nodes than to reach that state, unless the target state is stable \cite{Cornelius2011}.  Finally, we comment on the essential literature omitted in Ref.\ \cite{Wang2011}.\\

\medskip
\textbf{\emph{(i) Minimum number of driver nodes.}}
\bigskip

\noindent
Arguing that the control of a nonlinear system can be achieved by ``linearizing the nonlinear system about [a] finite number of local desired states,"
Ref.\ \cite{Wang2011} claims that the controllability of a network with nonlinear node dynamics can be determined by the structure of so-called ``local effective networks" (LENs). For one-dimensional nonlinear node dynamics, this leads to the conclusion that ``any bidirectional network system can be fully controlled 
by a single driver node, regardless of the network topology" \cite{Wang2011}.  While these claims are stated for rather general systems,  networks of $n$ diffusively coupled dynamical units of the following form are used as a model system throughout  Ref.\ \cite{Wang2011}:   
\begin{eqnarray}\label{eq:0}
\dot{x}_i &=& F_i(x_i) + \sum_{j=1}^{n}a_{ij}[H(x_j)-H(x_i)] + \sum_{k=1}^{m}b_{ik}u_k(t)\nonumber\\
             &=& F_i(x_i) - \sum_{j=1}^{n}\ell_{ij}H(x_j) + \sum_{k=1}^{m}b_{ik}u_k(t),
\end{eqnarray}
where $x=[x_i]_{n\times 1}$ and
$x_i$ stands for the state of node $i$, which we assume to be one-dimensional for convenience of presentation. Here, $F_i$ represents the intrinsic (nonlinear) dynamics of node $i$,  $H$ represents the coupling function, $A=[a_{ij}]_{n\times n}$ is the adjacency matrix of the network, $L=[\ell_{ij}]_{n\times n}$ is the Laplacian matrix of the network, and $B=[b_{ij}]_{n\times m}$ is the control matrix (i.e., $u_j(t)$ is the $j$th time-dependent control signal). Following Ref.~\cite{Wang2011}, a node $i$ in the network is a {\it driver node} if the node is directly controlled, that is,  there is some $k$ such that $b_{ik}$ is nonzero.  
(Because an irreducible set of $s$ independent input control signals can always be assigned one-to-one to $s$ different nodes even if other nodes in the network receive a combination of the same inputs, it can be argued that $s$ is more important than the number of driver nodes---number of nodes for which ``there are input signals on them"  \cite{Wang2011}. But we show below that the results in  Ref.\ \cite{Wang2011} do not hold when expressed in terms of $s$  either.)
 
System (\ref{eq:0}) is {\it controllable} if for {\it every} given initial state $x^{(0)}$ and target state $x^{(1)}$ in the state space  there exists a piecewise continuous input signal $u(t)=[u_j(t)]_{m\times 1}$ such that the system can be steered from $x^{(0)}$  to $x^{(1)}$ in a finite time. Therefore, the notion of controllability is a {\it global} property of the system for a {\it given} input matrix $B$. This property has a local counterpart, referred to as {\it local controllability}. Given a matrix $B$, the system is locally controllable at a given state $x^{(0)}$ if there is a neighborhood around this state such that every point in this neighborhood can be reached from $x^{(0)}$ by a control trajectory in a finite time.
The authors' identification of the minimum sets of driver nodes required to achieve full control of system~\eqref{eq:0} is based on the linearization of the dynamics.  However, we show that even within the specific class of systems explicitly considered in Ref.\ \cite{Wang2011}, the linearized dynamics cannot determine the minimum number of driver nodes, let alone their identities. 

\begin{figure}[t] 
\centering
\includegraphics*[width=0.9\textwidth]{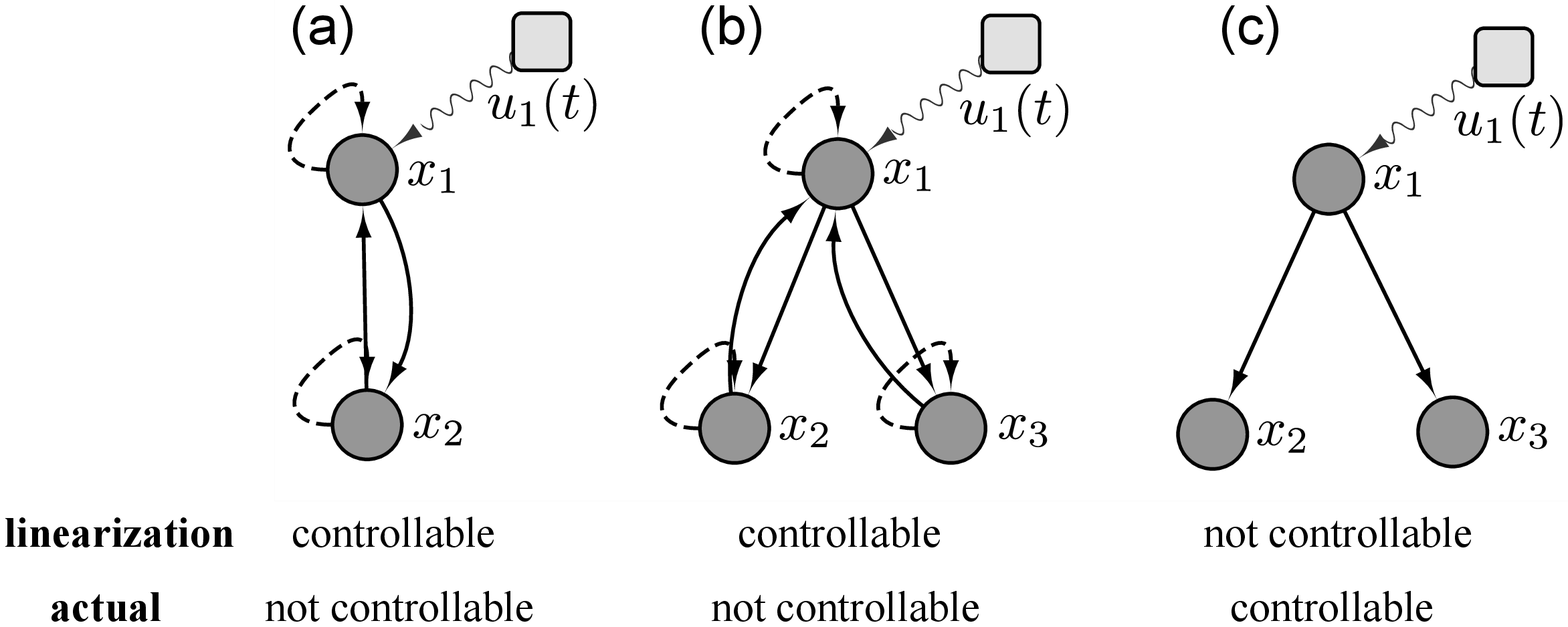}
\caption{\baselineskip 14pt
Counterexamples to the assertion that the controllability of a network with nonlinear dynamics can be predicted from the linearization procedure proposed in Ref.\ \cite{Wang2011}. The solid lines represent edges in the network, while dashed self-loops represent additional edges in the LENs derived from Ref.\ \cite{Wang2011}.
(a, b) Examples of networks whose linearized dynamics are controllable while their nonlinear counterparts are not: (a) bidirectional network with two nodes corresponding to the dynamics in Eq.\ (\ref{eq:10}) and (b) bidirectional network with three nodes corresponding to the dynamics in Eq.\ (\ref{eq:15}). 
(c) Example of a network whose nonlinear dynamics is controllable while its linearized counterpart is not:
star three-node network endowed with the dynamics $\dot{x}_1= u_1(t)$, $\dot{x}_2=a_{21}x_1$, $\dot{x}_3=a_{31}x_1^3$,
where $a_{21}\neq 0$ and $a_{31}\neq 0$.   The conclusion that this system is controllable follows from Theorem 2.1 in Ref.\ \cite{Sun2010}. Therefore, the ``controllability" determined by the linearization-based method of Ref.\ \cite{Wang2011} fails both as  a necessary and as a sufficient condition for the controllability of the actual nonlinear dynamics.}
\label{fig:0}
\end{figure}

Counterexamples to the claims in Ref.\ \cite{Wang2011} are abundant, even for networks with only two nodes. For instance, consider the system\vspace{-0.1cm}
\begin{equation}\label{eq:10}
\begin{cases}
	\dot{x}_1 = 2x_1^2 +  a_{12}(x_2^2-x_1^2) + b_1 u_1(t), \\
	\dot{x}_2 = x_2^2 + a_{21}(x_1^2-x_2^2) + b_2 u_2(t),
\end{cases}
\end{equation}
which corresponds to a two-node network in which $H(x)=x^2$ (see Fig.~\ref{fig:0}(a)).  System~\eqref{eq:10} is controllable if both $b_1$ and $b_2$ are  nonzero (i.e., all nodes are driven). The case of interest is the one in which only one of the $b_i$'s is nonzero (one driver node). For example, let $b_2=0$ (similar result holds for $b_1=0$). In this case,  $x_2$ cannot be decreased from {\it any} initial state in the state space if $a_{21}\in(0,1)$, and thus system~\eqref{eq:10} is neither globally nor locally controllable. Since the lack of controllability holds for all $a_{21}\in(0,1)$,  system~\eqref{eq:10} is {\it not} controllable with a single driver node, even though the network has only two nodes.  Note that one would indeed have reached the opposite conclusion by analyzing the associated LEN as proposed in Ref.\ \cite{Wang2011}.  Following Ref.\ \cite{Wang2011}, the linearized dynamics around an arbitrary  state $x^{(l)}=(x^{(l)}_1,x^{(l)}_2)^T$ is given by \vspace{-0.2cm}
\begin{equation}
 \begin{pmatrix}
\dot\xi_1 \\
\dot\xi_2\
 \end{pmatrix}
 =
 J\cdot \begin{pmatrix}
\xi_1 \\
\xi_2\
 \end{pmatrix}
 =
 \begin{pmatrix}
  2(2- a_{12})x^{(l)}_1 & 2 a_{12} x^{(l)}_2 \\
  2 a_{21}x^{(l)}_1 & 2(1- a_{21})x^{(l)}_2\
 \end{pmatrix} \cdot 
 \begin{pmatrix}
\xi_1 \\
\xi_2\
 \end{pmatrix},
\end{equation}
and the LEN is a network with edges corresponding to the nonzero entries of matrix $J$. For the range of parameters $0<a_{21}<1$ and $x^{(l)}_1\neq0$, for which we have just shown that system~\eqref{eq:10} is not even locally controllable, the corresponding LEN is a two-node network containing a directed edge from node $1$ to node $2$ and a self-loop at node $2$ (the other edges are irrelevant), and is thus predicted by the analysis of Ref.~\cite{Wang2011} to be structurally controllable through a single control signal input at node $1$. This contradiction invalidates the use of LEN for the analysis of controllability in networks of nonlinear dynamics. 

In fact, counterexamples are common even when the coupling functions are linear (e.g., $H(x)=x$). Consider a star network with three nodes, with the hub being the only node where the control signal is applied (Fig.~\ref{fig:0}(b)). 
The network dynamics is expressed as
\begin{equation}\label{eq:15}
\begin{cases}
	\dot{x}_1 = F_1(x_1) + a_{12}(x_2-x_1) + a_{13}(x_3-x_1) +u_1(t), \\
	\dot{x}_2 = x_2^2 +   a_{21}(x_1-x_2), \\
	\dot{x}_3 = 2x_3^2 + a_{31}(x_1-x_3),
\end{cases}
\end{equation}
which, according to the theory in Ref.~\cite{Wang2011}, is fully controllable for any node dynamics $F_1$ for almost all $a_{21}$ and $a_{31}$ such that $a_{21}a_{31}> 2(a_{21}-a_{31})^2$. 
However, all networks within this range of parameters are in fact not controllable and this follows from Theorems 2.1 and 2.2 in Ref.\ \cite{Sun2007a}. Therefore, the statement that ``the minimum number of driver nodes to achieve full control of the system is determined by the structural properties of the LENs" \cite{Wang2011} is false in general. As further elaborated below, this is so because the controllability of the linearized dynamics does {\it not} imply (even {\it locally}) the controllability of its nonlinear counterpart.

We also comment on the controllability of networks with nonlinear dynamics in general. As shown by Theorem 1 in Ref.\ \cite{Sun2007b}
and by Theorem 3.1 in Ref.\ \cite{Hunt1982}, a nonlinear dynamical system with $n$ state variables may not be controllable for any given number of independent input control signals smaller than $n$.  For example, system (\ref{eq:10}) with a single input control signal, $u_1(t)=u_2(t)$, is not controllable for any $a_{12}$, $a_{21}$, $b_1$, and $b_2$ in the range $[b_1a_{21}-(2-a_{12})b_2]\cdot[b_1(1-a_{21})-b_2a_{12}]>0$, even though both nodes are driven.  Therefore, this example also rules out the possibility of recasting the Ref.\ \cite{Wang2011}'s statement in terms of the number $s$ of input signals instead of the number of driver nodes, since it shows that bidirectional networks with one-dimensional nonlinear node dynamics are not guaranteed to be controllable by one input signal, either. On the other hand, the possibility of controlling a particular nonlinear system with a small number of inputs is not ruled out. The controllability of complex networks with nonlinear dynamics thus depends on both the network structure and the form of the dynamics, in sharp contrast to what is stated in Ref.~\cite{Wang2011}.  
\\

\medskip
 \textbf{\emph{(ii)  Failure of the control method.}}
\bigskip

\noindent
The control method proposed in Ref.\ \cite{Wang2011} is based on linearizing the nonlinear system about a finite number of local desired states. The reason this approach fails might appear intriguing, since the linearized dynamics approximates the nonlinear system in sufficiently small neighborhoods of each state. However,  when the linearized system  is driven between two states in one such neighborhood, the resulting trajectory is not necessarily limited to that neighborhood.  This is shown below for linear systems in general, which include as special cases the linearized dynamics considered in Ref.\ \cite{Wang2011}.  In fact, for such systems, even if the control trajectory to go from state $x^{(1)}$ to state $x^{(2)}$ is strictly local, in general the control trajectory to go from state $x^{(2)}$ to state $x^{(1)}$  will be necessarily nonlocal whenever the number of driver nodes is smaller than $n$.  The implication of this is clear:  the controllability of the linearized dynamics relies on ``global" control trajectories and hence this form of linearization cannot be reliably used to infer the controllability of the original nonlinear system, even locally.

To formalize this observation, consider the  linear control system in the form \vspace{-0.15cm}
\begin{equation}\label{eq:1}\vspace{-0.15cm}
\dot{x} = {\cal A}x + {\cal B}u(t),  
\end{equation}
where $x=[x_1,\dots,x_n]^T\in\mathbb{R}^n$ is the vector of state variables, $u=[u_1,\dots,u_m]^T$ is the vector of time-dependent control signals, and ${\cal A}=[{\cal A}_{ij}]_{n\times n}$ and ${\cal B}=[{\cal B}_{ij}]_{n\times m}$ are constant matrices (${\cal B}$ plays the role of the input matrix $B$ in system\ (\ref{eq:0})). 

\bigskip
\noindent 
{\it Definition 1.\ (Strictly local controllability)} - The linear system~\eqref{eq:1} is strictly locally controllable at $x^{(0)}$ if for any $\varepsilon>0$ there exists $\delta>0$ such that, for any target state $x^{(1)}$ satisfying $||x^{(1)}-x^{(0)}||<\delta$, the following property holds: there exists a piecewise continuous input signal $u(t)$ defined on a finite time interval $[t_0,t_1]$
under which  the solution of~\eqref{eq:1} satisfies $x(t_0)=x^{(0)}$, $x(t_1)=x^{(1)}$, and  $||x(t)-x^{(0)}||<\varepsilon$ for all $t_0\leq t\leq t_1$. On the other hand, the system is not strictly locally controllable at $x^{(0)}$ if there exists an $\varepsilon>0$, such that for any $\delta>0$ there is a target state $x^{(1)}$ with $||x^{(1)}-x^{(0)}||<\delta$ for which the following property holds: any control input that steers the system from $x^{(0)}$ to $x^{(1)}$ along a trajectory $x(t)$ for $t_0\leq t\leq t_1$ with $x(t_0)=x^{(0)}$ and $x(t_1)=x^{(1)}$ contains at least one point $x(\tau)$,  for $t_0<\tau<t_1$,  such that $||x(\tau)-x^{(0)}||>\varepsilon$. (Our definition of strictly local controllability is analogous to the concept of small-time local controllability often used in the context of nonlinear systems \cite{Kawski1990}.)

\begin{figure}[b!] 
\centering
\includegraphics*[width=0.62\textwidth]{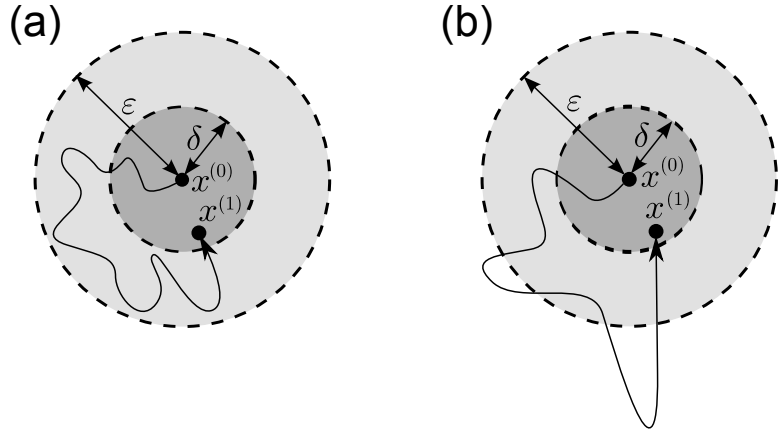}
\vspace{-0.3cm}
\caption{\baselineskip 14pt
Schematic representation of strictly local controllability. 
(a)  Illustration of a strictly locally controllable state $x^{(0)}$: nearby states can be reached from this state through a control trajectory that is limited to a small neighborhood.
(b)  Illustration of a locally controllable state $x^{(0)}$ that is {\it not} strictly locally controllable: there are nearby states that can only be reached through control trajectories that are not limited to a small neighborhood. 
}\label{fig:1}
\end{figure}

\bigskip
Figure~\ref{fig:1} illustrates this definition. Note that a state of the system can be locally controllable (i.e., it can be driven to any other state in a sufficiently small neighborhood) without being strictly locally controllable. A notable exception is when the state is a fixed point of the dynamics, as it can be proved that the fixed point of a linear system (which is unique and at the origin for nonsingular matrix ${\cal A}$) is always strictly locally controllable. On the other hand, unless all the nodes are taken as driver nodes, almost every other point in the state space is not strictly locally controllable. 

\bigskip
\noindent
{\it Proposition 1.\ (Breakdown of strict local controllability)} - 
Consider the linear control system~\eqref{eq:1}. If there is at least one component $x_k$ that is not directly controlled (i.e., ${\cal B}_{kj}=0$ for all $j$), then the system is not strictly locally controllable almost everywhere. 

\begin{figure}[b]
\centering       
\includegraphics*[width=0.45\textwidth]{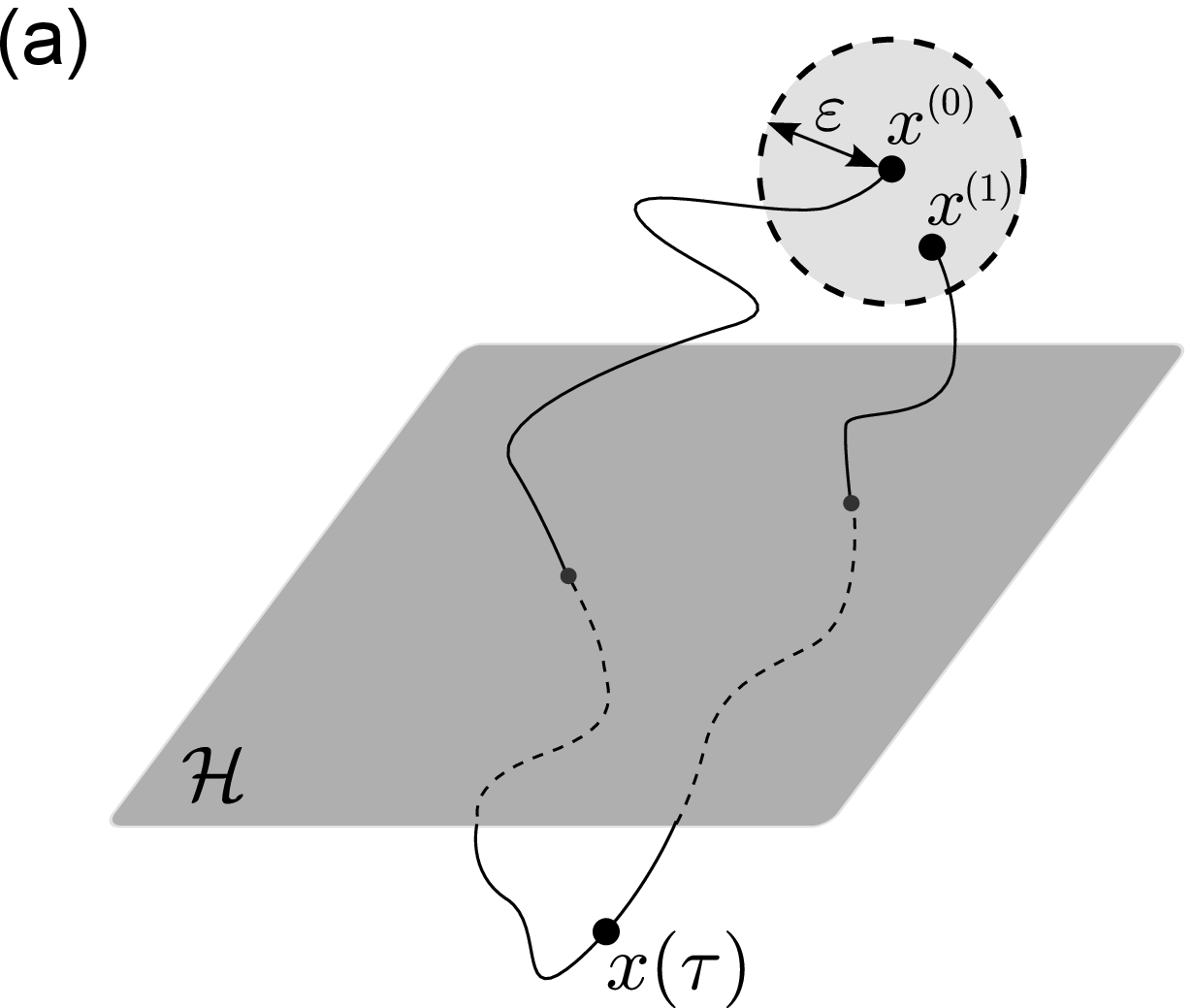}
\hspace{0.4cm}
\includegraphics*[width=0.51\textwidth]{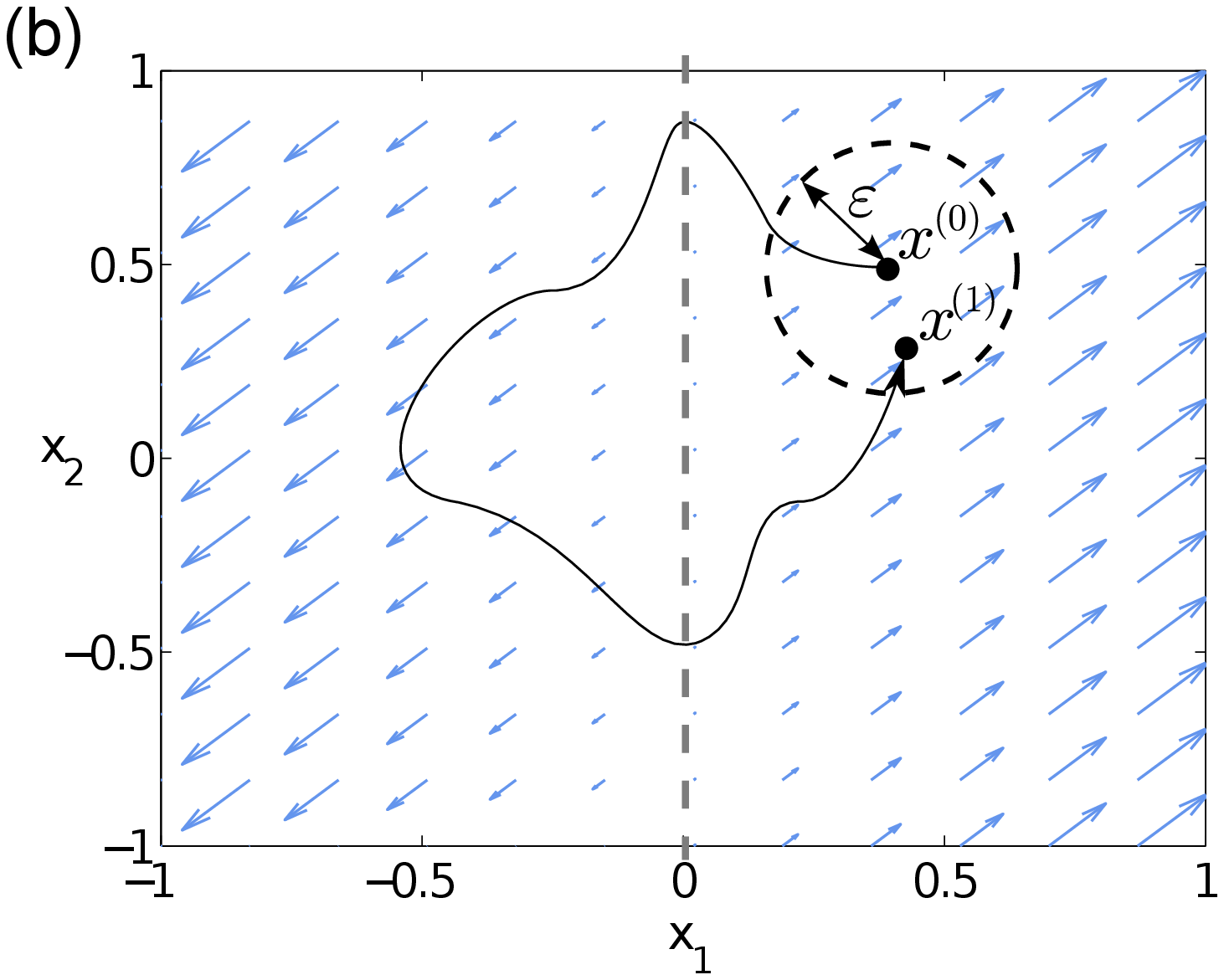}
\caption{\baselineskip 14pt
Illustration that almost all states of linear systems are not strictly locally controllable.
(a) Schematic representation of  this property as formalized in the proof of  Proposition 1.
(b) Example two-dimensional system $\dot{x}_1 = x_1 + u_1(t)$, $\dot{x}_2 = x_1$: any point not on the line $x_1=0$ is not strictly locally controllable. The background arrows indicate the vector field in the absence of 
control. The curve represents a typical control trajectory to bring state $x^{(0)}$ to state $x^{(1)}$, illustrating that
whenever these states are in the $x_1>0$ half of the plane and  $x^{(1)}_2<x^{(0)}_2$, the control trajectory has to visit the  $x_1<0$ half of the plane. Note that this is so because not all the variables are directly controlled, which is tantamount to not controlling all the nodes in a network.
}\label{fig:2}
\end{figure}

\bigskip
\noindent
{\em Proof.} If ${\cal A}_{kj}=0$ for all $j$, the system is clearly not controllable.  If ${\cal A}_{kj}\neq0$ for some $j$, we
define the $(n-1)$-dimensional hyperplane $\mathcal{H}=\{x:({\cal A}x)_k=0\}$, where we have used the notation $({\cal A}x)_k=\sum_{j=1}^{n}{\cal A}_{kj}x_j$. The hyperplane $\mathcal{H}$ separates the state space $\mathbb{R}^n$ into two disjoint regions: $\mathcal{H}^+=\{x:({\cal A}x)_k>0\}$ and $\mathcal{H}^-=\{x:({\cal A}x)_k<0\}$.  Let the initial state be $x^{(0)}\in\mathcal{H}^+$. Choose $\varepsilon>0$ such that $B_{\varepsilon}(x^{(0)})\cap\mathcal{H}=\emptyset$, where $B_{\varepsilon}(x^{(0)})$ denotes the (open) ball centered at $x^{(0)}$ with radius $\varepsilon$. This also implies $B_{\varepsilon}(x^{(0)})\cap\mathcal{H}^-=\emptyset$. For any $\delta>0$, choose a target state $x^{(1)}$ defined by $ x^{(1)}_i=x^{(0)}_i  \mbox{~if $i\neq{k}$}$ and  $x^{(1)}_i=x^{(0)}_i - \delta/2  \mbox{~if $i=k$}$. It follows that $||x^{(1)}-x^{(0)}||=\delta/2<\delta$. In order for the trajectory $x(t)$ with $x(t_0)=x^{(0)}$ to reach $x^{(1)}$ in finite time (i.e., $\exists\, t_1<\infty$, such that $x(t_1)=x^{(1)}$), it is necessary that $x(\tau)\in\mathcal{H}^-$ for some $t_0<\tau<t_1$, since otherwise $\dot{x}_k = ({\cal A}x)_k$ would be nonnegative and $x_k$ would not decrease. Since $B_{\varepsilon}(x^{(0)})\cap\mathcal{H}^-=\emptyset$ by definition of $\varepsilon$, we have $x(\tau)\notin B_{\varepsilon}(x^{(0)})$. See Fig.~\ref{fig:2}(a) as an illustration. Similarly, we can show that any $x\in\mathcal{H}^-$ is not strictly locally controllable.  Since $\mathbb{R}^n=\mathcal{H}^{+}\cup\mathcal{H}^{-}\cup\mathcal{H}$ and $\mathcal{H}$ is a set of measure zero, the proposition follows.
\hfill{$\Box$}

\bigskip
Figure \ref{fig:2}(b) illustrates this general result for a network consisting of two nodes with one-dimensional intrinsic dynamics. With the exception of the line $x_1=0$, all the states of this system are not strictly locally controllable if node $1$ is taken as the only driver node.

Now, let us return to the specifics of what is done in  Ref.\ \cite{Wang2011}.  They write the variational equation representing the linearization of the nonlinear network dynamics (\ref{eq:0}) around a local desired state $x^{(l)}$  (Eq.~(1)  in \cite{Wang2011}) in the form \vspace{-0.1cm}
\begin{equation} \vspace{-0.1cm} \label{ref1:1}
\dot \xi = [DF(x^{(l)})-L\otimes DH(x^{(l)})]\cdot \xi,  
\end{equation}
where $\xi$ denotes the deviation vector such that $\xi(t=0)=x^{(0)}-x^{(l)}$  for the system at the  local initial state $x^{(0)}$ at $t=0$. They then go on to write the control problem  (Eq.~(2)  in \cite{Wang2011}) in the form \vspace{-0.15cm}
\begin{equation}  \vspace{-0.05cm} \label{ref1:2}
\dot \xi = G\xi +Bu(t), 
\end{equation}
where $G$ represents the matrix inside the brackets in Eq.\ (\ref{ref1:1}).
 Their approach is illustrated in Fig.\ 1 in Ref.\ \cite{Wang2011}---a figure strikingly similar to Fig.\ 1(b) in Ref.\ \cite{Cornelius2011}---and
is then summarized as follows:  ``Once the variational states $\xi$ are controlled to approach zero values, the original nonlinear coupled system moves to the local desired state. The next local desired state can then be chosen. The final desired state can be reached by repeating this process. Due to the strongly structural controllability at each step, the whole controlling process is also strongly structurally controllable." 

Before proceeding we note that Eq.\ (\ref{ref1:1}) is incorrect. If the linearization was really around the fixed state $x^{(l)}$ (as implied in their text), as opposed to the trajectory of $x^{(l)}$, then the resulting linearized equation would have the extra additive term $F(x^{(l)})-L\otimes H(x^{(l)})$. This term vanishes, in agreement with  Eq.\ (\ref{ref1:1}) (Eq.\ (1) in \cite{Wang2011}), only in the exceptional case when  $x^{(l)}$  is a fixed-point solution of the uncontrolled dynamics. In typical cases, when $x^{(l)}$  is not a fixed-point solution, the control of this linear system to reach $x^{(l)}$ is equivalent to the control of a homogeneous linear system away from the origin. For example, when $G$ is invertible, the coordinate transformation $\eta=\xi+[DF(x^{(l)})-L\otimes DH(x^{(l)})]^{-1}[F(x^{(l)})-L\otimes H(x^{(l)})]$ brings the problem to the form of Eq.\ (\ref{eq:1}) but with the target state at $[DF(x^{(l)})-L\otimes DH(x^{(l)})]^{-1}[F(x^{(l)})-L\otimes H(x^{(l)})]$. The proposed method would then fail because, as shown in Proposition 1, the system is not strictly  locally controllable at those points and hence the linear approximation is violated by the control trajectory (except for the very special cases in which all local desired states are fixed points). If instead we accept the authors' form of the linearized dynamics (i.e., with no additive term), then $\xi$ in Eq.\ (\ref{ref1:1}) must represent the time evolution of the deviation between the trajectories starting at $x^{(l)}$  and $x^{(0)}$, which means that the argument of  $DF$ and $DH$ should in fact be the time evolution of $x^{(l)}$ rather than $x^{(l)}$ itself. In Ref.\  \cite{Wang2011} it is claimed (as a crucial condition within their approach) that matrix $G$ is a constant, which, in the best case scenario, is only an approximation since $G$ varies along the trajectory of $x^{(l)}$.  This can be criticized in its own right, but let us assume for the moment that this approximation is acceptable.

One might then be tempted to reason that, because the approach in Ref.\ \cite{Wang2011} is based on a linearization around the origin (with respect to $\xi$),  the breakdown of strict local controllability would not come into play. However, unless $x^{(l)}$ is a fixed point of the uncontrolled dynamics, $\xi$ being controlled to zero in a finite time $T$ does not mean that the control trajectory will reach the desired state $x^{(l)}$---this only implies that the control trajectory will meet the uncontrolled trajectory of $x^{(l)}$ at time $T$.   While one can reduce the convergence time $T$ arbitrarily and hence bring the meeting point to be arbitrarily close to the desired state $x^{(l)}$ (at least within the framework of unconstrained control), in general this necessarily causes the control trajectory to go arbitrarily far away from the origin (except for the unlikely situation in which $\xi_k$ is zero for the components of all nodes that are not directly controlled, i.e., when they are already equal to those of the desired state). Therefore, the trajectory is necessarily nonlocal in all nontrivial cases and the linearization becomes an invalid approximation of the nonlinear dynamics.  This can be formalized as follows.

\bigskip
\noindent
{\it Proposition 2.\ (Time lengthening versus trajectory diversion to reach the origin)} -  Consider the linear control system~\eqref{eq:1}. Suppose that there is at least one component $x_k$ that is not directly controlled (i.e., ${\cal B}_{kj}=0$ for all $j$). Let $x^{(0)}=(x^{(0)}_1,\dots,x^{(0)}_n)$ be the initial state and $x^{(1)}=(0,\dots,0)$ be the target state. Let $x(t)$ be a control trajectory such that $x(0)=x^{(0)}$ and $x(T)=x^{(1)}$, and let $\varepsilon = \max_{0\leq t\leq T}||x(t)||$.  Then $T \ge |x^{(0)}_k|/(\sum_{j}|{\cal A}_{kj}|\varepsilon)$. 

\bigskip
\noindent
{\em Proof.}
Along any control trajectory $x(t)$ with $||x(t)||\le\varepsilon$, we have
$|\dot{x}_k(t)| = |\sum_{j}{\cal A}_{kj}x_j| \leq \sum_{j}|{\cal A}_{kj}|\max_{j}|x_j| \leq \sum_{j}|{\cal A}_{kj}|\varepsilon$.
Since $x_k(0)=x^{(0)}_k$ and $x_k(T)=0$, it follows that
 $|x^{(0)}_k| \leq T \sum_{j}|{\cal A}_{kj}|\varepsilon$.
\hfill $\Box$

\bigskip
In the case of a network with one-dimensional node dynamics, having at least one component $x_k$ that is not directly controlled corresponds to having a number of driver nodes smaller than $n$ (the case of a single driver node is, of course, the most extreme of all such cases). Therefore, as anticipated above, it follows immediately from Proposition~2 that in general any attempt to make $T$ small will  lead $\varepsilon =\max_{0\leq t\leq T}||x(t)||$ to be very large. When $\varepsilon$ is large, the problem becomes nonlocal and linear approximation fails, as in the case of  non-strictly local controllability established in Proposition 1.\\

\medskip
\textbf{\emph{(iii) Failure of variants of the control method.}}
\bigskip

\noindent
The presentation in Ref.\ \cite{Wang2011} involves a number of contradictions. These include the inconsistent definition and use of $\xi$ and the use of $x^l$ (represented by $x^{(l)}$ here) to mean incompatibly different objects in the text, equations and figures (sometimes different things in different parts of the same equation). It is thus instructive to clarify that alternative interpretations of the authors' purported method are also destined to failure.  

Specifically, since $x^{(0)}$ cannot in general be brought to a given point $x^{(l)}$ with the linearization procedure proposed in Ref.\ \cite{Wang2011}, one may ask whether steering the trajectory of $x^{(0)}$ towards a different point of the trajectory of  $x^{(l)}$ would be any more promising. Needless to say, if control can only bring the trajectory of $x^{(0)}$ to meet the trajectory of $x^{(l)}$ at a point in the forward part of the trajectory with respect to $x^{(l)}$ itself, then the system cannot be controlled in a direction that goes against the flow (except for very special cases). The case in which the meeting point would be in the backward part of the trajectory of $x^{(l)}$ is potentially more promising because then the system could evolve spontaneously to the desired state $x^{(l)}$. However, we now show that, unless the desired state is in a small neighborhood of a fixed point of the original (nonlinear) dynamics, this is generally not possible for most initial states without violating the linear approximation.

\bigskip
\noindent
{\it Proposition 3.\ (Local drift of uncontrolled variables)} - 
Consider the control system in the form $\dot{x}={\cal F}(x) + {\cal B}u(t)$, where ${\cal F}$ can be nonlinear, and assume that $x_k$ is not directly controlled (i.e., ${\cal B}_{kj}=0$ for all $j$). Let   $x^{(l)}$ be a state such that ${\cal F}(x)_k>0$ for all $x$ in a neighborhood $B_{\varepsilon}(x^{(l)})$ for some $\varepsilon >0$. A state $x^{(1)} \in B_{\varepsilon}(x^{(l)})$ cannot be reached from any initial state $x^{(0)} \in B_{\varepsilon}(x^{(l)})$ with $x^{(0)}_k\geq x^{(1)}_k$ through a control trajectory inside $B_{\varepsilon}(x^{(l)})$.

\bigskip
\noindent
{\em Proof.} By assumption, $\dot{x}_k={\cal F}(x)_k>0$ for all $x\in B_{\varepsilon}(x^{(l)})$,
regardless of the input $u(t)$. 
It follows that any trajectory $x(t)$ inside $B_{\varepsilon}(x^{(l)})$ with
$x(t_0)=x^{(0)}$ 
necessarily satisfies $x_k(t)>x^{(0)}_k\geq x^{(1)}_k$ for all $t>t_0$.
\hfill{$\Box$}

\bigskip

This proposition implies that, in general, for most initial states $x^{(0)}$ in the neighborhood of $x^{(l)}$, 
one cannot reach any point $x^{(1)}$ in the backward trajectory of  $x^{(l)}$ by means of a local control 
trajectory.  This property underlies the breakdown of strictly local controllability in both linear and nonlinear systems. 
\\

\medskip
 \textbf{\emph{(iv) Linear versus nonlinear systems.}}
\bigskip

\noindent
It is claimed in  Ref.\ \cite{Wang2011} that
``network systems with nonlinear dynamics are more controllable than with linear dynamics."
This conclusion is based on the authors' interpretation that for networks with linear dynamics 
``there are limitation[s] on the network controllability, depending on the particular network structure,"
while networks with (one-dimensional) nonlinear dynamics can be ``fully controllable by one
controller at any node, regardless of the network structure." We have already demonstrated that their conclusion 
about networks with nonlinear dynamics is incorrect. We now show that their interpretation of the controllability 
of networks  with linear dynamics is also incorrect.

The authors' criterion is entirely based on the minimum number of driver nodes to control the network.  According to this criterion, however,  networks with nonlinear dynamics cannot be more controllable than those with linear dynamics for the class of systems considered in Ref.\ \cite{Wang2011}. This is so simply because connected bidirectional networks with one-dimensional linear dynamics are always controllable by a single driver node for almost all weights of the edges whenever the intrinsic node dynamics is non-degenerate, as expected in general. This has been clearly demonstrated in Ref.\ \cite{Cowan2011v1}, where the case of directed networks was also properly considered, and can also be anticipated from Ref.\ \cite{Lin1974}. Within the  model system (\ref{eq:0}), considered in Ref.\ \cite{Wang2011},  the linear case corresponds to  taking $F_i(x_i)=c_ix_i$ and $H(x_i)= x_i$, where $c_i$ is a constant,  and the non-degeneracy condition corresponds to at most one $c_i$ being zero (this assures that  $p_i\equiv c_i-\sum_{j\neq i}a_{ij}$ are all distinct for almost all choices of edge weights). The results of Ref.\  \cite{Cowan2011v1} are in fact strikingly similar to those of Ref.\  \cite{Wang2011}, except for the fact that the former are prior and correct while the latter are not. Incidentally,  despite all this, Ref.\  \cite{Cowan2011v1} is not referenced in Ref.\  \cite{Wang2011}.

Having shown that, according to the criterion of Ref.\ \cite{Wang2011}, networks with nonlinear dynamics could not be ``significantly more  controllable" than networks with linear dynamics, we clarify that for the purpose of this comparison their criterion itself is  misleading. Control concerns not only bringing the network to a specific state, but also keeping it at that state. Keeping the network at a state other than a stable state generally requires direct control of all nodes (a known fact also discussed in Ref.\ \cite{Cowan2011v3}). This is illustrated in Fig.\ \ref{fig:2}(b), where the system can be brought from $x^{(0)}$ to  $x^{(1)}$ by controlling only one of the two nodes but the direct control of both nodes is required to keep the system at  $x^{(1)}$. This is a very general and widely appreciated property in control theory.

While the linearization procedure and the concept of LEN proposed in Ref.\  \cite{Wang2011} do not contribute to our understanding of the controllability of nonlinear systems,  other valid inroads into this area had already been announced and are ignored in Wang {\it et al.}'s account of the literature.  An obvious precedent ignored in Ref.\  \cite{Wang2011}  is Ref.\ \cite{Cornelius2011}, where some of us developed a general method to control networks with nonlinear dynamics under rather general conditions (including arbitrary constraints on the control interventions, which are expected to be crucial in realistic situations).  The method in Ref.\ \cite{Cornelius2011} drives the system to the basin of attraction of a desired stable state (from which the uncontrolled system can evolve spontaneously to this desired state), and hence does not depend on continuous control to keep the system in that  state. The existence of multiple stable states, including desirable ones, is possible due to the nonlinear nature of the dynamics (nonsingular linear systems have at most one stable state) and is known to be common in real complex networks of interest, such as food webs, metabolic networks, and power grids.  Within this framework, it was already noted in Ref.\ \cite{Cornelius2011}  that ``The possibility of directing a complex network to a predefined dynamical state [...] can be achieved under rather general conditions by systematically designing compensatory perturbations that [...] take advantage of the full basin of attraction of the desired state, thus capitalizing on (rather than being obstructed by) the nonlinear nature of the dynamics."   This is one criterion (accounting for the fact that the system has to be kept at the desired state) in which networks with nonlinear dynamics are indeed more ``controllable" than networks with linear dynamics \cite{Cornelius2011}.  Another angle from which the same conclusion has been reached is when the control itself is allowed to be nonlinear, owing to independent directions of controlled motion (in addition to those from the linearized control dynamics) generated by the Lie algebra of the control vector fields \cite{Nijmeijer1990,Jurdjevic1996,Sontag1998,Lynch2005}.

The algorithm presented in Ref.\ \cite{Cornelius2011} to design control interventions is based on iterative applications of the variational equation to identify state perturbations that bring the closest approach point of the trajectory incrementally closer to the desired state. Equation (1) in Ref.\ \cite{Wang2011}, on which their method is built, is an ill-conceived implementation of the same variational equation to a situation where, as explained above, it cannot be used. 
Therefore, the whole framework in Ref.\ \cite{Wang2011} has major problems that go well beyond the missing literature and which render their results  unusable.
A main reason iterative linearization can be used in Ref.\ \cite{Cornelius2011}'s formulation  (but not in Ref.\ \cite{Wang2011}'s)  is that to reach a target basin of attraction all it is needed is that any point of the basin be accessible from the initial state. The identification of such points is a challenging problem due to the constraints and the high dimensionality typical of large real networks, and this is a central question addressed in Ref.\ \cite{Cornelius2011}. \\

\medskip
\textbf{\emph{(v) Final remarks.}}
\bigskip

\noindent
In summary, the results in the paper of W.-X.\ Wang, Y.-C.\ Lai, J.\ Ren,  B.\ Li \& C.\ Grebogi~\cite{Wang2011} are invalid. While we have focused on networks of one-dimensional dynamical systems, which form the bulk part of their work, it is immediate that their generalizations to high-dimensional node dynamics and networks with directed edges are also invalid. 
Unfortunately, the control method proposed in Ref.\  \cite{Wang2011} cannot be easily fixed  because, as shown here, the results of their linearization-based method and local effective networks fail both as a necessary and as a sufficient condition for the controllability of the actual nonlinear dynamics (Fig.\ \ref{fig:0}).  

Note that most of the analysis presented in this comment (in the form of simple but rigorous propositions) is expressed in a much stronger form than needed to point out the flaws in Ref.\  \cite{Wang2011}.  
For example, our analysis reveals comprehensive limits for the controllability of networks with nonlinear dynamics based on the proposed local linearization for any number of driver nodes smaller than $n$ (the case of a single driver node is, of course, the worst of all such  cases).
In particular,  the present arguments show that, based on local linearization, even for $n-1$ driver nodes such networks would be far less controllable than claimed in Ref.~\cite{Wang2011} for a single driver node. 

Leaving aside the technical problems, the concept of controllability that is the focus of  Ref.\  \cite{Wang2011}  concerns the ability of steering the system from any given state to all other states in the state space. In practice, however, this notion is often of limited usefulness since what is often needed (in both engineered and natural networks) is control within specific regions of the state space. In addition, far more important than the ability to move the system between any two states in the state space is the ability to account for constraints on the controls that are unavoidable in realistic situations.
These constraints often take the form of limiting the direction and magnitude of the control at certain nodes,  while forbidding the direct control of others.   Positive advances along these lines can and have been made \cite{Cornelius2011,Sahasrabudhe2011}.  We hope the limitations of the approach commented upon here  will not discourage others from contributing to this important area of research.\\


\vspace{2cm}

\vspace{2cm}

\begin{thebibliography}{99}
\baselineskip 16pt

\bibitem{Wang2011}
W.-X.~Wang, Y.-C.~Lai, J.~Ren, B.~Li, and C.~Grebogi,
Controllability of complex networks with nonlinear dynamics,
arXiv:1107.2177v1, posted on 12 July 2011.

\bibitem{Cornelius2011}
S.~P.~Cornelius, W.~L.~Kath, and A.~E.~Motter,
Controlling complex networks with compensatory perturbations,
arXiv:1105.3726v1,  posted on 18 May 2011.

\bibitem{Cowan2011v1}
N.~J.~Cowan, E.~J.~Chastain, D.~A.~Vilhena, and C.~T.~Bergstrom,  
Controllability of real networks,
arXiv:1106.2573v1, posted on 13 June 2011;  [with J.~S.~Freudenberg] arXiv:1106.2573v2, posted on 25 June 2011.

\bibitem{Cowan2011v3}
N.~J.~Cowan, E.~J.~Chastain, D.~A.~Vilhena, J.~S.~Freudenberg, and C.~T.~Bergstrom,
Nodal dynamics determine the controllability of complex networks,
arXiv:1106.2573v3, posted on 19 July 2011.

\bibitem{Haynes1970}
G.~W.~Haynes and H.~Hermes, 
Nonlinear controllability via Lie theory,
SIAM J. Control {\bf 8}, 450 (1970).

\bibitem{Sussmann1972}
H.~J.~Sussmann and V.~Jurdjevic,
Controllability of nonlinear systems,
J. Differ. Equations {\bf 12}, 95 (1972).

\bibitem{Hermann1977}
R.~Hermann and A.~J.~Krener,
Nonlinear controllability and observability,
IEEE Trans. Automat. Control {\bf 22}, 728 (1977).

\bibitem{Sun2007a}
Y.~Sun,
Necessary and sufficient condition for global controllability of planar affine nonlinear systems,
IEEE Trans. Autom. Control {\bf 52}, 1454 (2007).

\bibitem{Sun2007b}
Y.~Sun, S.~Mei, and Q.~Lu,
Necessary and sufficient condition for global controllability of a class of affine nonlinear systems,
J. Syst. Sci. Complex. {\bf 20}, 492 (2007).

\bibitem{Hunt1982}
L.~R.~Hunt,
$n$-Dimensional controllability with ($n-1$) controls,
IEEE Trans. Autom. Control. {\bf 27}, 113 (1982). 

\bibitem{Kawski1990}  
M.~Kawski,
High-order small-time local controllability, 
in: {\it Nonlinear Controllability and Optimal Control}, edited by H. J. Sussmann 
(Marcel Dekker, New York, NY, 1990), pp.\ 431--467.

\bibitem{Lin1974}
C.-T.~Lin,
Structural controllability,
IEEE Trans. Autom. Control. {\bf 19}, 201 (1974).

\bibitem{Nijmeijer1990}
H.~Nijmeijer and A.~van~der~Schaft,
{\it Nonlinear Dynamical Control Systems}
(Springer-Verlag, New York, NY, 1990).

\bibitem{Sontag1998}
E.~Sontag,
{\it Mathematical Control Theory: Deterministic Finite Dimensional Systems}
(2nd ed., Springer-Verlag, New York, NY, 1998).

\bibitem{Jurdjevic1996}
V.~Jurdjevic,
{\it Geometric Control Theory}
(Cambridge University Press, Cambridge, UK, 1997).

\bibitem{Lynch2005}
K.~Lynch,
Nonholonomic and underactuated systems,
in: {\it Principles of Robot Motion: Theory, Algorithms, and Implementations} 
(The MIT Press, Cambridge, MA, 2005),  pp.\ 401--472.

\bibitem{Sun2010}
Y. Sun,
Global controllability of a class of 3-dimensional affine nonlinear systems,
Proceedings of the 29th Chinese Control Conference July 29--31, 2010, Beijing, China.

\bibitem{Sahasrabudhe2011}
S. Sahasrabudhe and A. E. Motter,
Rescuing ecosystems from extinction cascades through compensatory perturbations, 
Nat. Commun. {\bf 2}, 170 (2011). 

\end{thebibliography}
\end{document}